\documentclass[conference,10pt]{IEEEtran}
\IEEEoverridecommandlockouts

\usepackage{epsfig,rotating,setspace,latexsym,amsmath,epsf,amssymb,amsfonts,bm,theorem,cite,algorithm,graphicx,epsf,authblk,epstopdf,color,algpseudocode,bbm}

\newtheorem{lemma}{Lemma}
\newenvironment{Proof}[1]{\medskip\par\noindent{\bf Proof:\,}\,#1}{{\mbox{\,$\blacksquare$}\par}}

\allowdisplaybreaks

\begin{document}

\title{Age Minimization in Energy Harvesting Communications: Energy-Controlled Delays\thanks{This work was supported by NSF Grants CNS 13-14733, CCF 14-22111, CCF 14-22129, and CNS 15-26608.}}

\author{Ahmed Arafa \quad Sennur Ulukus\\
\normalsize Department of Electrical and Computer Engineering\\
\normalsize University of Maryland, College Park, MD 20742\\
\normalsize \emph{arafa@umd.edu} \quad \emph{ulukus@umd.edu}}

\maketitle

\begin{abstract}
We consider an energy harvesting source that is collecting measurements from a physical phenomenon and sending updates to a destination within a communication session time. Updates incur transmission delays that are function of the energy used in their transmission. The more transmission energy used per update, the faster it reaches the destination. The goal is to transmit updates in a {\it timely} manner, namely, such that the total {\it age of information} is minimized by the end of the communication session, subject to energy causality constraints. We consider two variations of this problem. In the first setting, the source controls the number of measurement updates, their transmission times, and the amounts of energy used in their transmission (which govern their delays, or service times, incurred). In the second setting, measurement updates externally arrive over time, and therefore the number of updates becomes fixed, at the expense of adding data causality constraints to the problem. We characterize age-minimal policies in the two settings, and discuss the relationship of the age of information metric to other metrics used in the energy harvesting literature.
\end{abstract}

\section{Introduction}

A source collects measurements from a physical phenomenon and sends information updates to a destination. The source relies solely on energy harvested from nature to communicate, and the goal is to send these updates in a {\it timely} manner during a given communication session time, namely, such that the total {\it age of information} is minimized by the end of the session time. The age of information is the time elapsed since the freshest update has reached the destination.

Power scheduling in energy harvesting communication systems has been extensively studied in the recent literature. Earlier works \cite{jingP2P, kayaEmax, omurFade, ruiZhangEH} consider the single-user setting under different battery capacity assumptions, with and without fading. References \cite{jingBC, omurBC, jingMAC, kaya-interference} extend this to multiuser settings: broadcast, multiple access, and interference channels; and \cite{ruiZhangRelay, gunduz2hop, berkDiamond-jour, varan_twc_jour, arafa_baknina_twc_dec_proc} consider two-hop, relay, and two-way channels.

Minimizing the age of information metric has been studied mostly in a queuing-theoretic framework; \cite{yates_age_1} studies a source-destination link under random and deterministic service times. This is extended to multiple sources in \cite{yates_age_mac}. References \cite{ephremides_age_random, ephremides_age_management, ephremides_age_non_linear} consider variations of the single source system, such as randomly arriving updates, update management and control, and nonlinear age metrics, while \cite{shroff_age_multi_hop} shows that last-come-first-serve policies are optimal in multi-hop networks.

Our work is most closely related to \cite{yates_age_eh, elif_age_eh}, where age minimization in single-user energy harvesting systems is considered; the difference of these works from energy harvesting literature in \cite{jingP2P, kayaEmax, omurFade, ruiZhangEH, jingBC, omurBC, jingMAC, kaya-interference, ruiZhangRelay, gunduz2hop, berkDiamond-jour, varan_twc_jour, arafa_baknina_twc_dec_proc} is that the objective is age of information as opposed to throughput or transmission completion time, and the difference of them from age minimization literature in \cite{yates_age_1, yates_age_mac, ephremides_age_random, ephremides_age_management, ephremides_age_non_linear, shroff_age_mdp, shroff_age_multi_hop} is that sending updates incurs energy expenditure where energy becomes available intermittently. \cite{yates_age_eh} considers random service time (time for the update to take effect) and \cite{elif_age_eh} considers zero service time. Recently in \cite{arafa-age-2hop}, we considered a fixed non-zero service time in two-hop and single hop settings. In our work here, we consider an energy-controlled (variable) service time in a single-user setting.

We consider a source-destination pair where the source relies on energy harvested from nature to send information updates to the destination. Different from \cite{yates_age_eh, elif_age_eh}, updates' service times depend on the amounts of energy used to send them; the higher the energy used to send an update, the faster it reaches the destination. Hence, a tradeoff arises; given an amount of energy available at the source, it can either send a few number of updates with relatively small service times, or it can send a larger number of updates with relatively higher service times. In this paper, we investigate this tradeoff and characterize the optimal solution in the offline setting. We formulate the most general setting of this problem where the source decides on the number of updates to be sent, when to send them, and the amounts of energy consumed in their transmission (and therefore the amounts of service times or delays they incur), such that the total age of information is minimized by the end of the session time, subject to energy causality constraints. We present some structural insights of the optimal solution in this general setting, and propose an iterative solution. Our results show that the optimal number of updates depends on the parameters of the problem: the amounts and times of the harvested energy, delay-energy consumption relationship, and the session time.

We also consider the scenario where update arrival times at the source (measurement times) cannot be controlled; they arrive during the communication session. Thus, two main changes occur to the previously mentioned model. First, the total number of updates gets fixed; and second, data causality constraints are enforced, since the source cannot transmit an update before receiving it. We formulate the problem in this setting and characterize its optimal solution.

\section{System Model and Problem Formulation}

A source node acquires measurement updates from some physical phenomenon and sends them to a destination during a communication session of duration $T$ time units. Updates need to be sent as {\it timely} as possible, i.e., such that the total {\it age of information} is minimized by time $T$. The age of information metric is defined as
\begin{align}
a(t)\triangleq t-U(t),\quad\forall t
\end{align}
where $U(t)$ is the time stamp of the latest received information (measurement) update, i.e., the time at which it was acquired at the source. Without loss of generality, we assume $a(0)=0$. The objective is to minimize the following quantity
\begin{align}
A_T\triangleq\int_{0}^Ta(t)dt
\end{align}

\begin{figure}[t]
\center
\includegraphics[scale=.9]{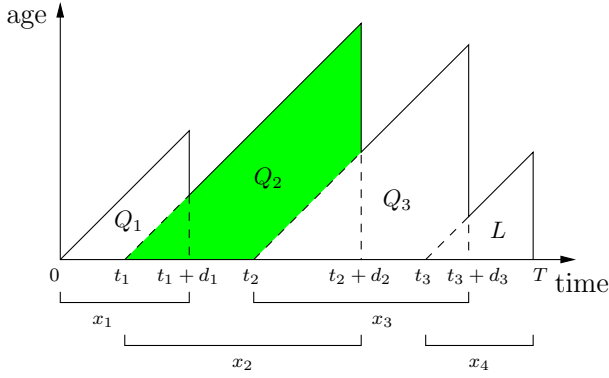}
\caption{Age evolution versus time in a controlled measurement times system, with $N=3$ updates.}
\label{fig_age_ex_ctrl}
\end{figure}

The source powers itself using energy harvested from nature, and is equipped with an infinite battery to store its incoming energy. Energy is harvested in packets of sizes $E_j$ at times $s_j$, $1\leq j\leq M$. Without loss of generality, we assume $s_1=0$. The total energy harvested by time $t$ is
\begin{align} \label{eq_tot_en}
\mathcal{E}(t)=\sum_{j:~s_j\leq t}E_j
\end{align}
We denote by $e_i$, the energy used in transmitting update $i$, and denote by $d_i$, its transmission delay (service time) until it reaches the destination. These are related as follows
\begin{align}
e_i=f(d_i)
\end{align}
where $f$ is a deceasing convex function\footnote{This relationship is valid, for instance, if the channel is AWGN. With normalized bandwidth and noise variance, we have $f(d)=d\left(2^{2B/d}-1\right)$, with $B$ denoting the size of the update packet in bits \cite{cover}.}. Let $t_i$ denote the transmission time of update $i$. The following then holds
\begin{align} \label{eq_en_caus}
\sum_{i=1}^kf(d_i)\leq\sum_{i=1}^k\mathcal{E}(t_i),\quad\forall k
\end{align}
which represent the {\it energy causality constraints} \cite{jingP2P}, which mean that energy cannot be used in transmission prior to being harvested. We also have the {\it service time constraints}
\begin{align} \label{eq_srv}
t_i+d_i\leq t_{i+1},\quad\forall i
\end{align}
which ensure that there can be only one transmission at a time.

\subsection{Controlled Measurements}

In this setting, the source controls when to take a new measurement update, and the goal is to choose total number of updates $N$, transmission times $\{t_i\}_{i=1}^N$, and delays $\{d_i\}_{i=1}^N$, such that $A_T$ is minimized, subject to energy causality constraints in (\ref{eq_en_caus}) and service time constraints in (\ref{eq_srv}). We note that the source should start the transmission of an update measurement whenever it is acquired. Otherwise, its age can only increase. In Fig.~\ref{fig_age_ex_ctrl}, an example run of the age evolution versus time is presented in a system with $N=3$ updates. The area under the age curve is given by the sum of the areas of the three trapezoids $Q_1$, $Q_2$, and $Q_3$, plus the area of the triangle $L$. The area of $Q_2$ for instance is given by $\frac{1}{2}\left(t_2+d_2-t_1\right)^2-\frac{1}{2}d_2^2$. Computing the area for a general $N$ updates, we formulate the problem as follows
\begin{align} \label{opt_gen}
\min_{N,{\bm t},{\bm d}}\quad&\sum_{i=1}^N\left(t_i+d_i-t_{i-1}\right)^2-d_i^2+\left(T-t_N\right)^2\nonumber\\
\mbox{s.t.}\quad &t_i+d_i\leq t_{i+1},\quad1\leq i\leq N\nonumber\\
&\sum_{i=1}^kf(d_i)\leq\sum_{i=1}^k\mathcal{E}(t_i),\quad1\leq k\leq N
\end{align}
with $t_0\triangleq0$ and $t_{N+1}\triangleq T$.

\begin{figure}[t]
\center
\includegraphics[scale=.9]{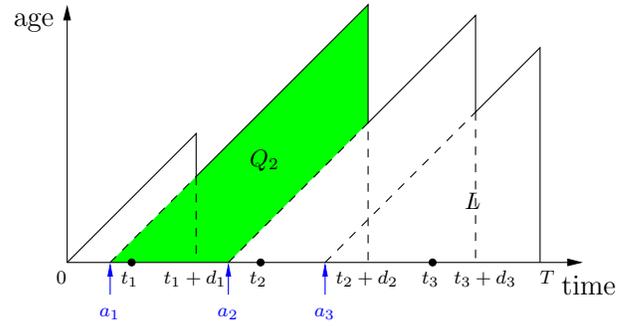}
\caption{Age evolution versus time in a system where $N=3$ update measurements arriving during communication.}
\label{fig_age_ex_arr}
\end{figure}

\subsection{Externally Arriving Measurements} \label{sec_sys_arr}

In this setting, measurement updates arrive during the communication session at times $\{a_i\}_{i=1}^N$, where $N$ is now fixed. We now have the following constraints
\begin{align}
t_i\geq a_i,\quad\forall i
\end{align}
representing the {\it data causality constraints} \cite{jingP2P}, which mean that updates cannot be transmitted prior to being received at the source. In Fig.~\ref{fig_age_ex_arr}, we show an example of the age evolution in a system with $N=3$ arriving updates. The area of $Q_2$ in this case is given by $\frac{1}{2}(t_2+d_2-a_1)^2-\frac{1}{2}(t_2+d_2-a_2)^2$ and the area of $L$ is the constant term $\frac{1}{2}(T-a_3)^2$. Computing the area for general $N$ update arrivals, we write the objective function as $\sum_{i=1}^N\left(t_i+d_i-a_{i-1}\right)^2-\left(t_i+d_i-a_i\right)^2$, with $a_0\triangleq0$. This can be further simplified after some algebra to get the following problem formulation\footnote{An inherent assumption in this model is that $0<a_1<a_2<\dots<a_N$. Otherwise the areas of the trapezoids become 0 and the problem becomes degenerate. Also, the parameters of the problem are such that it is feasible.}
\begin{align} \label{opt_arr}
\min_{{\bm t},{\bm d}}\quad&\sum_{i=1}^N\left(a_i-a_{i-1}\right)\left(t_i+d_i\right)\nonumber\\
\mbox{s.t.}\quad &t_i+d_i\leq t_{i+1},\quad1\leq i\leq N\nonumber\\
&\sum_{i=1}^kf(d_i)\leq\sum_{i=1}^k\mathcal{E}(t_i),\quad1\leq k\leq N\nonumber\\
&t_i\geq a_i\quad1\leq i\leq N
\end{align}

We note that both problems (\ref{opt_gen}) and (\ref{opt_arr}) are non-convex. One main reason is that the total energy arriving up to time $t$, $\mathcal{E}(t)$, is not concave in $t$. Henceforth, in the next sections, we solve the two problems when all the energy packets arrive at the beginning of communication, i.e., when $M=1$ energy arrival. In this case $\mathcal{E}(t)=E$, $\forall t$. The solutions in the case of multiple energy arrivals follow similar structures.

\section{Controlled Measurements}

In this section, we focus on problem (\ref{opt_gen}) with a single energy arrival. We first have the following lemma.

\begin{lemma} \label{thm_gen_en}
In problem (\ref{opt_gen}), all energy is consumed by the end of communication.
\end{lemma}

\begin{Proof}
By direct first derivatives, we observe that the objective function is increasing in $\{d_i\}$. Thus, if not all energy is consumed, then one can simply use the remaining amount to decrease the last service time and achieve lower age.
\end{Proof}

Next, we apply the change of variables $x_1\triangleq t_1+d_1$, $x_i\triangleq t_i+d_i-t_{i-1}$ for $2\leq i\leq N$, and $x_{N+1}\triangleq T-t_N$. Then, we must have $\sum_{i=1}^{N+1}x_i=T+\sum_{i=1}^Nd_i$, which reflects the dependence relationship between the variables. This can also be seen geometrically in Fig.~\ref{fig_age_ex_ctrl}. Then, the problem becomes
\begin{align} \label{opt_gen_x}
\min_{N,{\bm x},{\bm d}}\quad&\sum_{i=1}^{N+1}x_i^2-\sum_{i=1}^Nd_i^2\nonumber\\
\mbox{s.t.}\quad &x_1\geq d_1\nonumber\\
&x_i\geq d_i+d_{i-1},\quad2\leq i\leq N \nonumber\\
&x_{N+1}\geq d_N \nonumber\\
&\sum_{i=1}^{N+1}x_i=T+\sum_{i=1}^Nd_i\nonumber\\
&\sum_{i=1}^Nf(d_i)\leq E
\end{align}

The variables $\{x_i\}_{i=1}^{N+1}$ control the inter-update times, which are lower bounded by the service times $\{d_i\}_{i=1}^N$, which are in turn controlled by the amount of harvested energy, $E$. We propose an iterative algorithm to find the optimal inter-update times $\{x_i^*\}$ given the optimal number of updates $N^*$ and the optimal service times $\{d_i^*\}$. This is described as follows.

Let $\{\bar{x}_i\}_{i=1}^{N+1}$ denote the output of this algorithm, and let us define the {\it stopping condition} to be when $\sum_{i=1}^{N^*+1}\bar{x}_i=T+\sum_{i=1}^{N^*}d_i^*$. We initialize by setting $\bar{x}_1=d_1^*$; $\bar{x}_i=d_i^*+d_{i-1}^*$, $2\leq i\leq N$; and $\bar{x}_{N+1}=d_N^*$. We then check the stopping condition. If it is not satisfied, we compute $m_1\triangleq\arg\min \bar{x}_i$, and increase $\bar{x}_{m_1}$ until either the stopping condition is satisfied, or $\bar{x}_{m_1}$ is equal to $\min_{i\neq m_1}\bar{x}_i$. In the latter case, we compute $m_2\triangleq\arg\min_{i\neq m_1}\bar{x}_i$, and increase both $\bar{x}_{m_1}$ and $\bar{x}_{m_2}$ {\it simultaneously} until either the stopping condition is satisfied, or they are both equal to $\min_{i\notin\{m_1,m_2\}}\bar{x}_i$. In the latter case, we compute $m_3$ and proceed similarly as above until the stopping condition is satisfied. Note that if $m_k$ is not unique at some stage $k$ of the algorithm, we increase the whole set $\{\bar{x}_i,~i\in m_k\}$ simultaneously.

The above algorithm has a water-filling flavor; it evens out the $x_i$'s to the extent allowed by the service times $d_i$'s and the session time $T$, while keeping them as low as possible. The next lemma shows its optimality.

\begin{lemma} \label{thm_gen_bar_algo}
In problem (\ref{opt_gen_x}), given $N^*$ and $\{d_i^*\}_{i=1}^N$, the optimal $x_i^*=\bar{x}_i$, $1\leq i\leq N+1$. 
\end{lemma}

\begin{Proof}
First, note that the algorithm initializes $x_i$'s by their least possible values. If this satisfies the stopping (feasibility) condition, then it is optimal. Otherwise, since we need to increase at least one of the $x_i$'s, the algorithm chooses the least one; this gives the least objective function since $y<z$ implies $(y+\epsilon)^2<(z+\epsilon)^2$ for $y,z\geq0$ and $\epsilon>0$. Next, observe that while increasing one of the $x_i$'s, if the stopping condition is satisfied, then we have reached the minimal feasible solution. Otherwise, if two $x_i$'s become equal, then by convexity of the square function, it is optimal to increase both of them simultaneously \cite{boyd}. This shows that each step of the algorithm is optimal, and hence it achieves the age-minimal solution.
\end{Proof}

We note that the above algorithm is essentially a variation of the solution of the single-hop problem in \cite{arafa-age-2hop}. There, all the inter-update delays are fixed, while here they can be different.

\begin{figure}[t]
\center
\includegraphics[scale=.45]{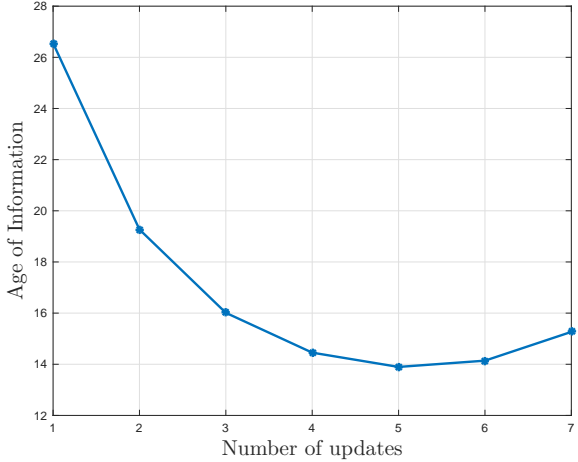}
\caption{Age of information versus number of updates.}
\label{fig_ex_ctrl_arr}
\end{figure}

Next, we present an example to show how the choice of the number of updates and inter-update delays affect the solution, in a specific scenario. In particular, we focus on the case where the inter-update delays are fixed for all update packets, i.e., $d_i=d,~\forall i$. In this case, by Lemma~\ref{thm_gen_en}, for a given $N$, the optimal inter-update delay is given by $d=f^{-1}\left(E/N\right)$. We can then use the algorithm above to find the optimal $x_i$'s, as shown in Lemma~\ref{thm_gen_bar_algo}. For example, we consider a system with energy $E=20$ energy units, with $f(d)=d\left(2^{2/d}-1\right)$, and $T=10$ time units. We plot the optimal age in this case versus $N$ in Fig.~\ref{fig_ex_ctrl_arr}. We see that the optimal number of updates is equal to $5$; it is not optimal to send too few or too many updates (the maximum feasible is $7$ in this example). This echoes the early results in \cite{yates_age_1}, where the optimal rate of updating is not the maximum (throughput-wise) or the minimum (delay-wise), but rather lies in between.

\section{Externally Arriving Measurements}

In this section, we solve problem (\ref{opt_arr}) with a single energy arrival.
We observe that the problem in this case is convex and can be solved by standard techniques \cite{boyd}. We first have the following lemma.

\begin{lemma}
In problem (\ref{opt_arr}), the optimal update times satisfy
\begin{align}
t_1^*&=a_1\\
t_i^*&=\max\left\{a_i,a_{i-1}+d_{i-1}^*,\dots,a_1+\sum_{j=1}^{i-1}d_j^*\right\},~i\geq 2
\end{align}
\end{lemma}

\begin{Proof}
This follows directly from the constraints of problem (\ref{opt_arr}); the optimal update times should always be equal to their lower bounds. Hence, we have $t_1^*=a_1$, $t_2^*=\max\{a_2,t_1^*+d_1^*\}=\max\{a_2,a_1+d_1^*\}$, $t_3^*=\max\{a_3,t_2^*+d_2^*\}=\max\{a_3,a_2+d_2^*,a_1+d_1^*+d_2^*\}$, and so on.
\end{Proof}

By the previous lemma, the problem now reduces to finding the optimal inter-update delays $\{d_i^*\}$. We note that starting from $t_1^*=a_1$, we have two choices for $t_2^*$; either $a_2$ or $a_1+d_1^*$. Once $t_2^*$ is fixed, $t_3^*$ in turn has two choices; either $a_3$ or $t_2^*+d_2^*$. Now observe that once a choice pattern is fixed, the objective function of problem (\ref{opt_arr}) will be given by $\sum_{i=1}^Nc_id_i$ where $c_i>0$ is a constant that depends on the choice pattern. For instance, for $N=3$, choosing the pattern $t_2^*=a_1+d_1^*$ and $t_3^*=a_3$ gives $c_1=a_2$, $c_2=a_2-a_1$, and $c_3=a_3-a_2$. We introduce the following Lagrangian for this problem \cite{boyd}
\begin{align}
\mathcal{L}=\sum_{i=1}^Nc_id_i+\lambda\left(\sum_{i=1}^Nf(d_i)-E\right)
\end{align}
where $\lambda$ is a non-negative Lagrange multiplier. The KKT conditions are
\begin{align}
c_i=-\lambda f^\prime(d_i)\label{eq_arr_kkt_d}
\end{align}
Hence, the optimal $\lambda^*$ is given by the unique solution of
\begin{align} \label{eq_arr_lmda}
\sum_{i=1}^Nh\left(-c_i/\lambda^*\right)=E
\end{align}
where $h\triangleq f\circ g$ and $g\triangleq\left(f^\prime\right)^{-1}$. To see this, note that since $f$ is convex, it follows that $g$ exists and is increasing. By (\ref{eq_arr_kkt_d}), we then have $d_i^*=g\left(-c_i/\lambda^*\right)$. Substituting in the energy constraint, which has to be satisfied with equality, gives (\ref{eq_arr_lmda}). By monotonicity of $f$ and $g$, $h$ is also monotone, and therefore (\ref{eq_arr_lmda}) has a unique solution in $\lambda^*$.

Therefore, we solve problem (\ref{opt_arr}) by first fixing a choice pattern for the update times, which gives us a set of constants $\{c_i\}$ allowing us to solve for $\lambda^*$ using (\ref{eq_arr_lmda}). We go through all possible choice patterns and choose the one that is feasible and gives minimal age.

We finally note that the measurements' arrival times can be so close to each other that the optimal solution is such that  $t_i^*>a_{i+l}$ for some $i$ and $l\geq1$. That is, there would be $l+1$ measurements waiting in the data queue before $t_i^*$. If the total number of updates can be changed, then this solution can be made better by transmitting only the freshest, i.e., the $(i+l)$th, measurement packet at $t_i^*$ and {\it ignoring} all the rest. This strictly improves the age and saves some energy as well. The solution can be further optimized by re-solving the problem with $\tilde{N}=N-l$ arriving measurements at times $\tilde{a}_1=a_1,\dots,\tilde{a}_{i-1}=a_{i-1},\tilde{a}_i=a_{i+l},\dots,\tilde{a}_{\tilde{N}}=a_N$.

\section{Discussion: Relationship to Other Metrics}

In this section we discuss the relationship between the proposed problems in this work and other well-known problems in the energy harvesting literature: transmission completion time minimization, and delay minimization. Reference \cite{jingP2P} introduced the transmission completion time minimization problem. In this problem, given some amounts of data arriving during the communication session, the objective is to minimize the time by which all the data is delivered to the destination, subject to energy and data causality constraints.

Reference \cite{tianDelay} studies this problem from a different perspective. Instead of minimizing the completion time of all the data, the objective is to minimize the delay experienced by each bit, which is equal to the difference between the time of its reception at the receiver and the time of its arrival at the transmitter. Delay-minimal policies are fundamentally different than those minimizing completion time. For instance, in \cite{jingP2P}, due to the concave rate-power relationship, transmitting with constant powers in between energy harvests is optimal. While in \cite{tianDelay}, the optimal delay-minimal powers are decreasing over time in between energy harvests, since earlier arriving bits contribute more to the cumulative delay and are thus given higher priorities (transmission powers and rates).

We note that minimizing the age of information problem is similar to the delay minimization problem formulated in \cite{tianDelay}. In both problems, there is a time counter that counts time between data transmissions and receptions. In the age of information problem, the time counter starts increasing from the beginning of the communication session. While in the delay problem, the time counter is bit-dependent; it starts increasing only from the moment a new bit enters the system and stops when it reaches the destination.

The delay minimization problem was previously formulated in \cite{jing-delay} for the case where the delay is computed per packet, as opposed to per bit in \cite{tianDelay} (note that the age is also computed per packet and not per bit). The transmitter in \cite{jing-delay} was energy constrained but not harvesting energy over time, which models the case where all energy packets arrive at the beginning of communication. For the sake of comparison, we extend the delay minimization problem in \cite{jing-delay} to the energy harvesting case as in \cite{tianDelay} and relate it to the age minimization problem considered in this work.

\begin{figure}[t]
\center
\includegraphics[scale=.9]{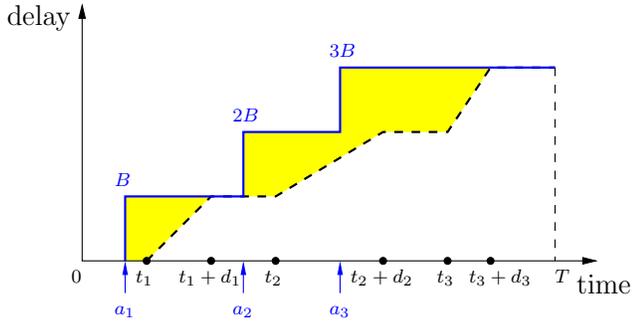}
\caption{Cumulative data packets arriving (blue) and departing (black) versus time with $N=3$ data packets. The shaded  area in yellow between the two curves represents the total delay of the system.}
\label{fig_delay_ex}
\vspace{-.2in}
\end{figure}

Following the model in Section~\ref{sec_sys_arr}, the $i$th arriving data packet waits for $t_i-a_i$ time in queue, and then gets served in $d_i$ time units. Following \cite{tianDelay}, the total delay is defined as the area in between the cumulative departing data curve, and the cumulative arriving data curve. In Fig.~\ref{fig_delay_ex}, we show an example realization using the same transmission, arrival, and service times used in Fig.~\ref{fig_age_ex_arr}. The solid blue curve represents the cumulative received data packets over time; the dotted black curve represents cumulative departed (served) data packets over time; and the shaded area in yellow represents the total delay $D_T$. The delay of the first data packet for instance is given by $B(t_1-a_1)+\frac{1}{2}Bd_1$, where $B$ is the length of the data packet in bits. Computing the area for general $N$ arrivals, the delay minimization problem is given by
\begin{align} \label{opt_delay}
\min_{{\bm t},{\bm d}}\quad&\sum_{i=1}^N2t_i+d_i\nonumber\\
\mbox{s.t.}\quad&\mbox{problem (\ref{opt_arr}) constraints}
\end{align}

We see that minimizing delay in problem (\ref{opt_delay}) is almost the same as minimizing age in problem (\ref{opt_arr}). The main difference is that to minimize age, transmission and service times are weighted by arrival times, while this is not the case when minimizing delay. The reason lies in the definitions of age and delay; the delay of a packet arriving at time $a$ stays the same if it arrives at time $a+\delta$, provided that its service time is the same, and that its transmission time {\it relative to its arrival time} is the same. The age of a packet on the other hand is directly affected by changing its arrival time as it represents the time stamp of when the packet arrived, and hence transmission and service times need to change if arrival times change in order to achieve the same age.

\end{document}